\documentclass[12pt]{article}
\usepackage{amsmath,amssymb,amscd,cite}
\newtheorem{thm}{Theorem}[section]
\newtheorem{prop}[thm]{Proposition}
\newtheorem{lem}[thm]{Lemma}
\newtheorem{cor}[thm]{Corollary}

\newtheorem{defi}[thm]{Definition}
\newcommand{\pf}{{\bf Proof. \ }}
\newcommand{\qed}{\hfill $\Box$ \\}

\font\msbm=msbm10 at 12pt
\newcommand{\Z}{\mbox{\msbm Z}}

\newcommand{\F}{\mbox{\msbm F}}
\newtheorem{rem}[thm]{Remark}

\date{}
\begin{document}
\title{Equivalence of  Quasi-cyclic Codes over Finite Fields}
\author{Kenza Guenda and T. Aaron Gulliver
\thanks{K. Guenda is with the Faculty of Mathematics USTHB, University
of Science and Technology of Algiers, Algeria. T. Aaron Gulliver
is with the Department of Electrical and Computer Engineering, University
of Victoria, PO Box 1700, STN CSC, Victoria, BC, Canada V8W 2Y2.
email: agullive@ece.uvic.ca. This work is an extension version of a part of Ph.D thesis of K. Guenda}}
\maketitle

\begin{abstract}
This paper considers the equivalence problem for quasi-cyclic codes over finite fields.
The results obtained are used to construct isodual quasi-cyclic codes.
\end{abstract}

\section{Introduction}

The equivalence problem for codes has numerous practical applications such as code-based cryptography~\cite{McEliece,otmani,sendrier3}.
As a consequence, many researchers have considered this problem~\cite{babai,job,sendrier1,sendrier3},
but to date there has been little progress in obtaining a solution.
Brand~\cite{brand} characterized the set of permutations by which two combinatorial cyclic objects on $p^r$ elements are equivalent.
Using these results, Huffman et al.~\cite{job} explicitly gave this set in the case $n=p^2$ and
provided algorithms to determine the equivalence between cyclic objects and extended cyclic objects.
In~\cite{job}, a negative answer was given to the generalization of their results to the case $n=p^r$, $r>2$.
Babai et al.~\cite{babai} gave an exponential time algorithm for determining the equivalence of codes.
Sendrier~\cite{sendrier1} proposed the support splitting algorithm to solve the problem of code equivalence in the binary case.
Unfortunately, in~\cite{sendrier3} it was shown that extending this algorithm to $q \geq5$ has an exponential growth in complexity.

In this paper, the equivalence problem is studied for quasi-cyclic codes over finite fields.
Tt is proven that two quasi-cyclic codes are equivalent if and only if their constituent codes are equivalent.
This is an important result which allows conditions to be given on the existence of isodual quasi-cyclic codes.
These conditions are used to obtain constructions of isodual quasi-cyclic codes.

The remainder of this paper is organized as follows.
In Section 2, some preliminary definitions and results are given.
The main result is presented in Section 3.
It is proven that two quasi-cyclic codes are equivalent if and only if their constituent codes are equivalent.
In Section 4, we introduce multiplier equivalent cyclic codes.
Further, we examine the equivalence of quasi-cyclic codes with cyclic constituent codes.
Section 5 then considers conditions on the existence of isodual quasi-cyclic codes.

\section{Preliminaries}

Let $C$ be a linear code of length $n$ over a finite field
$\mathbb{F}_q$, and $\sigma$ a permutation of the symmetric group
$S_n$ acting on $\{0,1,\ldots, n-1\}$.
We associate with this code a linear code $\sigma(C)$ defined by
\[
\sigma(C)=\{ (x_{\sigma ^{-1}(0)}, \ldots, x_{\sigma ^{-1}(n-1)});\, (x_0, \ldots x_{n-1})\in C \}.
\]
We say that the codes $C$ and $C'$ are equivalent if
there exists a permutation $\sigma \in S_n$ such that $C'=\sigma(C)$.
The automorphism group of $C$ is the subgroup of $S_n$ given by
\[
Aut(C)=\{\sigma \in S_n; \, \sigma(C)=C\}.
\]

A linear code $C$ of length $n$ over $\mathbb{F}_q$ is called
quasi-cyclic of index $l$ or an $l$-quasi-cyclic code if its automorphism group
contains the permutation $T^l$ given by
\begin{equation}
\begin{split}
T^l : \Z_n & \longrightarrow  \Z_n \\
          i &\longmapsto i +l \mod n.
\end{split}
\end{equation}
This definition is equivalent to saying that for all $c\in
{C}$ we have $T^l(c) \in {C}$ with $T : i \mapsto
i+1$ being the circular shift.
The index $l$ of ${C}$ is the smallest integer satisfying this property.
It can easily be proven that $l$ is a divisor of $n$.
If $l=1$ the code $C$ is called a cyclic code.
The automorphism group of $C$ then contains the cyclic shift $T$.
A cyclic code over $\F_q$ of length $n$ is an ideal of the ring $\F_q[x]/(x^n-1)$.
Hence it is generated by a polynomial $f(x) | (x^n-1)$.
For a primitive element $\alpha $ of $\F_q$, the defining set $T$ of a cyclic code is a subset of $\Z_n$; $T=\{i\le n, f(\alpha^i)=0\}$.
There is a one-to-one correspondence between the irreducible factors of $f(x)$ and subsets of $T$.
These subsets are called the cyclotomic classes.

Let $a$ and $n$ be positive integers such that $\gcd(a,n)=1$.
The permutation~$\mu_a$ defined on $\Z_n=\{0,1,\ldots,n\}$ by
\begin{equation}
\begin{array}{ccl}
\label{eq:ling}
\mu_a:\Z_n&\longrightarrow &
 \Z_n\\
 i&\mapsto & \mu_a(i) =ia,
 \end{array}
\end{equation}
is called a multiplier.
Multipliers play an essential role in code equivalence~\cite{G-G}.
We attach the standard inner product to $\F_{q}^n$
\[
[{v},{w}] = \sum v_iw_i.
\]
The Euclidean dual code $C^\perp$ of $C$ is defined as
\begin{equation}
C^\perp=\{ {v} \in \F_{q}^n; \ [{v},{w}]= 0 {\rm \  for\ all\ }
{w} \in C\}.
\end{equation}
If $C \subseteq C^\perp$, the code is said to be self-orthogonal, and
if $C=C^\perp$ the code is self-dual.
We call an isodual code a linear code which is equivalent to its dual.

Let $f(x)= a_0+a_1x+\ldots +a_rx^r$ be a polynomial of degree $r$
with $f(0)= a_0\neq 0$.
Then the monic reciprocal polynomial of $f(x)$ is
\[
f^*(x)= f(0)^{-1}x^rf(x^{-1}) = a_0^{-1}(a_r+a_{r-1}x+\ldots +a_0x^r).
\]
If a polynomial is equal to its reciprocal then it is called a self-reciprocal polynomial.

\section{Equivalent Quasi-cyclic Codes}

In this section, we characterize the equivalence problem for quasi-cyclic codes.

Let $\F_q$ be the finite field of cardinality $q$ and $m$ be a positive integer such that $\gcd(m,q)=1$.
Further, let $\F_{q}[Y]$ denote the ring of polynomials in the indeterminate $Y$ over $\F_q$.
Define the ring $R=\F_{q}[Y]/(Y^m-1)$, and for a positive integer $l$ define the following map
\begin{equation}
\label{eq:decom}
\begin{split}
\Phi :\F_{q}^{lm}  &  \longrightarrow R^l  \\
c=(c_{0,0},c_{0,1},\ldots, c_{0,l-1},\ldots,c_{r-1,0}, \ldots, c_{r-1,l-1}) &\longmapsto
 \Phi (c)= (c_0(Y),c_1(Y), \ldots, c_{l-1}(Y)),
\end{split}
\end{equation}
where $c_j(Y)= \sum \limits_{i=0}^{m-1}c_{i,j}Y^i \in \F_{q}$.
It was shown in~\cite{sole1} that the map $\Phi$ induce a one-to-one
correspondence between quasi-cyclic codes over $\F_{q}$ of index $l$ and
length $lm$ and linear codes over $R$ of length $l$.

Note that in~(\ref{eq:decom}) each coordinate $c_{i,j}$ in
$c=(c_{0,0},\ldots, c_{0,l-1},\ldots ,c_{m-1,0}, \ldots, c_{m-1,l-1})$
can be written as $c_{j+il}$, $0\le j\le l-1$, $1\le i \le m-1$.
Now $c_j(Y)=\sum \limits_{i=0}^{m-1}c_{ij}Y^i \in R$ can be expressed its vectorial form as
$c_j(Y)=(c_{0,j}, c_{1,j},\ldots, c_{m-1,j})$.
Then the image of the codeword $(c_{j+il})_{ 0\le j\le l-1;1\le i \le m-1}$
by the map $\Phi$ is the codeword $(c_{i+jm})_{ 0\le j\le l-1;1\le i \le m-1}$.
This suggests the following result.

\begin{prop}
\label{prop:image}
Let $\mathcal{C}$ and $\mathcal{C}'$ be quasi-cyclic codes of length $lm$ and index $l$ over $\F_q$.
Then $\mathcal{C}$ and $\mathcal{C}'$ are equivalent if and only if the codes
$C=\Phi(\mathcal{C})$ and $C'=\Phi(\mathcal{C})$ are equivalent.
\end{prop}
\pf
Assume that $\mathcal{C}=\{(c_{j+il})_{ 0\le j\le l-1;1\le i \le
m-1}\}$ and $\mathcal{C}'=\{(c'_{j+il})_{ 0\le j\le l-1;1\le i \le m-1}\}$
are equivalent by a permutation $\sigma \in S_n$.
Hence if $\sigma$ is such that $\sigma (j+il)= j'+i'l$,
then we have $\sigma((c_{j+il})_{0\le j\le l-1;1\le i \le m-1}) = (c'_{j+il})_{ 0\le j\le l-1;1\le i
\le m-1}= (c_{j'+i'l})_{ 0\le j'\le l-1;1\le i' \le m-1}$.
Hence
\[
\Phi(\sigma((c_{j+il})_{0\le j\le l-1;1\le i \le m-1})) =\Phi ((c_{j'+i'l})_{ 0\le j'\le l-1;1\le i' \le m-1})= (c_{i'+j'm})_{ 0\le j'\le l-1;1\le i' \le m-1},
\]
and we have an associated permutation $\tau$ given by $\tau(i'+j'm)=i+jm$.
Since $\sigma$ is in $S_n$, $\tau$ is also in $S_n$.
Furthermore, $\tau$ is such that $\tau(\Phi(\sigma(\mathcal{C}))=\Phi(\mathcal{C})$.
This proves the first implication.

Now assume that $C=\{(c_{i+jm})_{ 0\le j\le l-1;0\le i \le
m-1}\}$ and $C'=\{(c'_{i+jm})_{ 0\le j\le l-1;0\le i \le m-1}\}$ are
images by the map $\Phi$ of two quasi-cyclic codes $\mathcal{C}$ and
$\mathcal{C'}$, respectively,
and there exists a permutation $\sigma$ such that
$\sigma (C)=\sigma(\{(c_{i+jm})_{ 0\le j\le l-1;0\le i \le m-1}=C'=(c'_{i+jm})_{ 0\le j\le l-1;0\le i \le m-1}
\}=\{(c_{i'+j'm})_{0\le j\le l-1;0\le i \le m-1}\}$.
Hence $\mathcal{C}= \{(c_{j+il})_{0\le j\le l-1;0\le i \le m-1}\}$ and
$\mathcal{C'}= \{(c_{j'+i'l})_{0\le j'\le l-1;0\le i' \le m-1}\}$.
Then by defining the permutation
$\tau$ such that $\tau(j'+i'l)=j+il$ we obtain that
$\tau(\mathcal{C}')=\mathcal{C}$.
\qed

Now we consider the factorization of $Y^m-1$ over $\F_q$.
Since it is assumed that $\gcd (m,q)=1$, $Y^m-1$ has a unique decomposition into irreducible factors over $\F_q$
\begin{equation}
\label{eq:fact} Y^m-1=\delta g_1\ldots g_s h_1h_1^{*} \ldots h_t
h_t^*,
\end{equation}
where $\delta$ is a unit in $\F_{q}$, $h_i^{*}$ is the reciprocal of $h_i$, and $g_i$ is self-reciprocal.
The ring $R$ is a principal ideal ring, so it can be decomposed into a direct sum of local rings.
Hence the Chinese Remainder Theorem gives the following decomposition
\begin{equation}
\label{eq:decom2}
R=\frac{\F_{q}[Y]}{(Y^m-1)}=\left(\bigoplus_{i=1}^s
\frac{\F_{q}[Y]}{(g_i)}\right)\oplus\left(\bigoplus_{j=1}^t\left(\frac{\F_{q}[Y]}{(h_j)}\oplus
\frac{\F_{q}[Y]}{({h_j}^*)}\right)\right).
\end{equation}
Let $\frac{\F_{q}[Y]}{(g_i)}=G_i$, $\frac{\F_{q}[Y]}{(h_j)}=H_j'$, and $\frac{\F_{q}[Y]}{({h_j}^*)}=H_j''$.
Since the polynomials in the decomposition (\ref{eq:fact}) are irreducible,
the local rings are in fact field extensions of $\F_q$.
Then as a consequence of the decomposition (\ref{eq:decom2}), we obtain
that every $R$-linear code of length $l$ can be decomposed as
$C=(\oplus_{i=1}^s C_i )\oplus(\oplus_{j=1}^t (C_j'\oplus C_j''))$,
where $C_i$ is a linear code over $G_i$,
$C_j'$ is a linear code over $H_j'$,
and $C_j''$ is a linear code over $H_j''$.
The codes $C_i$, $C_j$ and $C_j''$
are called the {\it components} of the quasi-cyclic code $\mathcal{C}$.

Assume that $g_i$ is one of the self-reciprocal polynomials in~(\ref{eq:fact}).
We now study the action of the following map over the local component ring $\F_{q}[Y]/\langle g_i \rangle=G_j$ of $R$
\begin{equation}
\begin{array}{ccl}
\label{eq:ling} \,-\,  :\F_{q}[Y]/\langle g_i \rangle
&\longrightarrow &
\F_{q}[Y]/\langle g_i \rangle\\
c(Y)&\mapsto & c(Y^{-1})).
\end{array}
\end{equation}

The map $-$ is a ring automorphism.
For $g_i$ of degree 1 this map is the identity, and if $\deg(g_i)=K_i \neq 1$, since $g_i$ and $g_i^*$ are associated, $K_i$ must be even.
Since $g_i$ is irreducible and square free, it is also separable and local.
Further, as $g_i$ is irreducible of degree $d_i$,
from \cite[Theorem 4.2]{Mac2} the ring $G_i=\F_{q}[Y]/\langle g_i \rangle$ is an extension of $\F_q$,
namely $\F_{q^{d_i}}$.
Then the map $r\mapsto  \overline{r}$, is the map $\nu :r \mapsto r^{q^{K_i/2}}$ and is a power of the Frobenius map.
Hence, it is a permutation over $\F_{q^{d_i}}$ which fixes the elements of $\F_{q}$.
This proves the following result.

\begin{lem}
\label{lem:eqH}
With the previous notation, each code $C_i$ over $G_i$ is equivalent to $\nu(C_i)=\overline{C_i}$.
\end{lem}

For each $a=(a_0,\ldots,a_{l-1})$, $b=(b_0,\ldots,b_{l-1})$ in $G_{i}^l$, we define the Hermitian inner product on $G_i$ by
\begin{equation}
\label{eq:Herm}
\langle a,b \rangle^H= \sum _ka \overline{b_k}.
\end{equation}
This is in fact the usual Hermitian inner product.
We now have the following lemma.

\begin{lem}
\label{lem:mono}
Let $C_i$ be a linear code over $G_i$.
The Hermitian dual of $C_i$ denoted ${C_i}^{\bot H}$ is equivalent to the Euclidean dual of $C_i$.
\end{lem}
\pf
Define the code $\overline{C}=\{\overline{r};\, r\in C\}$. It is
easy to see that $ {C_i}^{\bot H}=\overline{(C_i)}^{\bot}=\nu (C_i)^{\bot}$.
Hence from Lemma~\ref{lem:eqH} we have that $\nu (C_i)^{\bot} = (\nu (C_i^{\bot})$.
\qed

For $a,b \in \F_q^{lm}$, let $\Phi(a)=(a_0,\ldots, a_{l-1})$ and $\Phi(b)=(b_0,\ldots, b_{l-1})$,
where
\[
a_i= (a_{i,1}, \ldots, a_{i,s},{a_{i,1}}', {a_{i,1}}'', \ldots, {a_{i,t}}'{a_{i,1}}''),
\]
and
\[
b_i= (b_{i,1}, \ldots, b_{i,s},{b_{i,1}}', {b_{i,1}}'', \ldots, {b_{i,t}}'{b_{i,1}}''),
\]
with $a_{i,j}, b_{i,j}\in G_j$,  ${a_{i,j}}', {b_{i,j}}'\in {H_j}'$, and ${a_{i,j}}'', {b_{i,j}}''\in {H_j}''$.

We define the Hermitian inner product on $R^l$ by
\begin{eqnarray}
\langle \Phi(a),\Phi(b) \rangle&=&\left( \sum_i {a_{i,1}} \overline{b_{i,1 }}, \ldots, \sum_i {a_{i,s}} \overline{b_{i,s }}, \right.
\nonumber \\
&&\sum_i {a_{i,1}}' {b_{i,1 }}'',  \sum_i {a_{i,1}}''{b_{i,1 }}', \ldots
\nonumber \\
&&\left. \sum_i {a_{i,t}}'{b_{i,t}}'',\sum_i {a_{i,t}}'' {b_{i,t }}'\right).\nonumber
\end{eqnarray}
Using this inner product, Ling and Sol\'e~\cite{sole1} and Lim~\cite{lim} gave the Euclidean dual of a quasi-cyclic code.
\begin{prop}
\label{propodual}
Let $\mathcal{C}$ be an $l$-quasi-cyclic code of length $lm$ over $\F_q$
and $C=\Phi(\mathcal{C})=(\oplus_{i=1}^s C_{i}\oplus(\oplus_{j=1}^t (C_{j}^{'}\oplus C_{j}^{''})))$ be its image as defined previously.
Then the Euclidean dual of $\mathcal{C}$ is the $l$-quasi-cyclic code $\mathcal{C}^{\bot}$ such that
$\Phi(\mathcal{C}^{\bot})=(\oplus_{i=1}^s C_{i}^{\bot H}\oplus(\oplus_{j=1}^t (C_{j}^{''\bot}\oplus C_{j}^{'{\bot}})))$.
\end{prop}

We require the following lemma concerning the direct sum of codes.
\begin{lem}
\label{lem:direct}
Assume that $C= C_1\oplus C_2$ and $C'= C_1'\oplus C_2'$ are codes of length $2n$ which are the direct
sums of codes of length $n$.
Then there exist a permutation $\sigma \in S_{2n}$ such that $\sigma(C) =C'$ if and only if there exists
permutations $\sigma_1$ and $\sigma_2$ in $S_n$ such that $\sigma_1(C_1)=C_1'$ and $\sigma_2(C_2)=C_2'$.
\end{lem}

\pf
Assume that
\[
\sigma(C)= \sigma(C_1\oplus C_2)= \mathcal{C}',
\]
and
\[
C'= C_1'\oplus C_2'=\{(c_{\sigma(1)},\ldots c_{\sigma(n)},c_{\sigma(n+1)},\ldots, c_{\sigma(2n)}),\mbox{ with }
(c_1,\ldots,c_n)\in C_1\mbox{ and }(c_{n+1},\ldots,c_{2n}) \in C_2 \}.
\]
This gives that $\sigma(i) \in\{1,\ldots,n\}$ for $1\le i\le n$, and
$\sigma(i) \in\{n+1,\ldots,2n\}$ for $n+1\le i\le 2n$.
Hence we can define the permutations $\sigma_1$ and $\sigma_2$ on $n$ elements by
$\sigma_1(1)=\sigma (1), \ldots, \sigma_1(n)= \sigma(n)$, and
$\sigma_2(1)=\sigma(n+1), \ldots, \sigma_2(n)=\sigma(2n)$.
Then $\sigma(C_1\oplus C_2)=\sigma_1(C_1)\oplus \sigma_2(C_2)= C_1'\oplus C_2'$.
Let the mapping $Pr_1$ be the projection on the first $n$ coordinates so that
$Pr_1( \sigma_1(C_1)\oplus \sigma_2(C_2))= \sigma_1(C_1)= Pr_1(C_1'\oplus C_2')=C_1'$ and
then $\sigma_1(C_1)=C_1'$.
We also obtain $\sigma_1(C_2)=C_2'$ by considering the projection $Pr_2$ on the last $n$ coordinates.
For the converse, assume that there exists permutations $\sigma_1$ and $\sigma_2$ such that
$\sigma_1(C_1)=C_1'$ and $\sigma_2(C_2))= C_2'$.
Hence we obtain the permutation $\sigma \in S_{2n}$ given by $\sigma(i)=\sigma_1(i)$,
and $\sigma(i+n)=\sigma_2(i)$ for $1 \le i \le n $, so then $\sigma(C)=C'$.
\qed

\begin{rem}
Lemma~\ref{lem:direct} is also true for the direct sum of $k>2$ codes of the same length.
\end{rem}

\begin{thm}
\label{main:thm}
Let $\mathcal{C}$ be a quasi-cyclic code of length $lm$ and index $l$ over $\F_q$ such that
$\Phi(\mathcal{C})=(\oplus_{i=1}^s C_i)\oplus(\oplus_{j=1}^t (C_j'\oplus C_j'')$.
Then $\mathcal{C}$ is isodual if and only if each of its components $C_i$ for $1\le i\le s$ is
isodual, and for each $1 \le j \le t$ we have that $C_j'$ is equivalent to $C_j''^{\bot}$.
\end{thm}
\pf
Let $\mathcal{C}$ be an $l$-quasi-cyclic code which is isodual.
Then there exists a permutation $\sigma$ such that $\mathcal{C}=\sigma (\mathcal{C}^{\bot})$.
By Proposition~\ref{prop:image}, there exists a permutation $\tau$ such that
$\Phi(\mathcal{C})=\tau(\Phi(\mathcal{C}^{\bot})$.
From~Proposition~\ref{propodual} we have that
$\Phi(\mathcal{C}^{\bot}) =\Phi(\mathcal{C})^{\bot H}=(\oplus_{i=1}^s( C_i^{\bot H})\oplus(\oplus_{j=1}^t
C_j''^{\bot}\oplus C_j'^{\bot}$.
Hence from Lemma~\ref{lem:direct} there exist permutations $\tau_i$, $\tau_j'$, and $\tau_j''$ such
that $C_i= \tau_i(C_i^{\perp H})$, $C_j'= \tau_j'(C_j'^\perp)$, and
$C_j'=\tau_j''(C_j''^\perp)$.
From Lemma~\ref{lem:mono} we have that ${C_i}^{\bot H}= \nu (C_i)^{\bot}$,
so $C_i= \tau_i(\nu(C_i^{\perp })$.
Then for $1\le i\le s$, the component $C_i$ is isodual.
For the converse, assume that each component of $\mathcal{C}$ is isodual.
Then we have that $\tau_i(C_i)=C_i^{\bot}$ for $1\le i\le s$,
$\tau_j'( C_j')=  C_j'^\bot$ and $\tau_j''(C_j'')=( C_j'')^{\bot}$ for $1\le j\le t$.
From Lemma~\ref{lem:mono} we have that $C_i^{\bot H}=\nu({C_i}^{\bot})$.
Hence $C_i^{\bot H}=\nu(\tau_i{C_i})$, so that
$\Phi(C)^{\bot }=(\oplus \nu(\tau_i( C_i)\oplus(\oplus \tau_j'( C_j')\oplus \tau_j''(C_j''))$.
Then from Lemma~\ref{lem:direct} there exists a permutation $\theta$ such that
$\Phi(C)^{\bot}= \theta (\oplus_{i=1}^s C_i)\oplus(\oplus_{j=1}^t (C_j'\oplus C_j''))$, and by
Proposition~\ref{prop:image} $\mathcal{C}$ is isodual.
\qed

The following corollary is a direct consequence of Proposition~\ref{prop:image} and Theorem~\ref{main:thm}.
Note that this result was given in \cite[Theorem 4.2]{sole1}.
\begin{cor}
\label{cor:condi}
An $l$-quasi-cyclic code $\mathcal{C}$ of length $lm$ over $R$ is self-dual if and only if
\[
\Phi(\mathcal{C})=\left(\bigoplus_{i=1}^s C_i\right)\bigoplus\left( \bigoplus_{j=1}^t \left(C_j'\bigoplus (C_j')^{\bot}\right)\right),
\]
where for $1\le i \le s$, $C_i$ is a self-dual code over $\frac{R[Y]}{(g_i)}$ with respect to the Hermitian inner product,
and for $1 \le j \le t$, $C_j'$ is a linear code of length $l$ over $H_j$ and $ C_j'^{\bot}$ is its dual with
respect to the Euclidean inner product.
\end{cor}

In~\cite[Proposition 6.1]{sole1},
conditions were given on the existence of self-dual quasi-cyclic codes of index $2$.
We generalize these results to give conditions on the existence of self-dual quasi-cyclic codes of index $l$ even as follows.

\begin{thm}
Let $m$ be an integer relatively prime to $q$.
Then self-dual quasi-cyclic codes over $\F_q$ of length $lm$, $l$ even, exists if
and only if one of the following conditions is satisfied:
\begin{enumerate}
\item[(i)] $q$ is a power of $2$,
\item[(ii)] $q=p^b$, where $p$ is a prime congruent to $1 \mod 4$, or
\item[(iii)] $q=p^{2b}$, where $p$ is a prime congruent to $3 \mod 4$.
\end{enumerate}
\end{thm}

\pf
If a self-dual quasi-cyclic code $\mathcal{C}$ over of length $lm$ exists, then Corollary~\ref{cor:condi}
shows that there is a self-dual code $C_1$ of length $l$ over $G_1$.
Hence the conditions in the theorem are necessary.
Conversely, if any one of the conditions is
satisfied, then there exists $\gamma \in \F_q$ such that $\gamma^2+1=0$.
Consequently, every finite extension of $\F_q$ also contains such an element.
Then the code generated by $(1,\gamma,\ldots, 1, \gamma)$ is self-dual over any extension of
$\F_q$ (with respect to both the Euclidean and Hermitian inner products).
Hence from Corollary~\ref{cor:condi}, a self-dual quasi-cyclic code of length $lm$ exists over $\F_q$.
\qed

\section{Multiplier Equivalent Quasi-Cyclic Codes}
A natural question that arises is, can a multiplier be a permutation by which two quasi-cyclic codes are equivalent?
In the special case of the so-called one-generator quasi-cyclic codes, Ling and Sol\'e defined the multiplier equivalence.
However, this definition can be placed in a more general setting than that given in~\cite{sole3},
namely there is no need to restrict the definition to one-generator quasi-cyclic codes.
From Lemma~\ref{lem:direct} and Proposition~\ref{propodual} we have that two quasi-cyclic codes are equivalent
if and only if their constituent codes are equivalent.
Hence we can give the following definition.
\begin{defi}
\label{defi:mutiplier}
Two quasi-cyclic codes $\mathcal{C}$ and $\mathcal{D}$ are multiplier equivalent if and only if each of their components
are multiplier equivalent.
\end{defi}

In the next section, conditions are given on when two quasi-cyclic codes with cyclic components are multiplier equivalent.

\subsection{Equivalence of Quasi-Cyclic Codes with Cyclic Constituent Codes}

In this section, we consider the equivalence of quasi-cyclic codes with cyclic constituent codes, i.e.
$\Phi(\mathcal{C})$ is cyclic or $\Phi(\mathcal{C})$ is an ideal of $R[X]/(X^l-1)$.
We have the following results.

\begin{prop}(\cite[Proposition 8]{lim})
\label{prop:lim}
Let $q$ be a prime power and $\F_q$ the finite field with $q$ elements.
Further, let $l$ and $m$ be positive integers with $m$ coprime to
$q$, and let $C$ be a quasi-cyclic code of length $lm$ and index $l$
over $\F_q$. Then the following are equivalent
\begin{enumerate}
\item[(i)] $\Phi(\mathcal{C})$ is cyclic, and
\item[(ii)] all the constituent codes of $\mathcal{C}$ are cyclic.
\end{enumerate}
\end{prop}

\begin{thm}
Let $\mathcal{C}$ and $\mathcal{D}$ be quasi-cyclic codes of length $pm$ and index $p$ a prime, both with cyclic constituent codes.
Then $\mathcal{C}$ and $\mathcal{D}$ are equivalent if and only if they are multiplier equivalent.
\end{thm}

\pf
Assume that $\mathcal{C}$ and $\mathcal{D}$ are quasi-cyclic codes with cyclic constituent codes.
Then from Proposition~\ref{prop:lim} all the constituent codes are cyclic.
Furthermore, from Theorem~\ref{main:thm} $\mathcal{C}$ and $\mathcal{D}$ are equivalent if and only if their
cyclic constituent codes are equivalent.
These cyclic codes have length $p$ a prime.
Then from~\cite[Theorem 1]{job}, they are equivalent if and only if they are multiplier equivalent. Hence the result follows.
\qed

\begin{rem}
When $l=p^{\alpha}, \alpha >1$, there exist other permutations by which two quasi-cyclic codes may be equivalent~\cite{G-G}.
\end{rem}

\begin{thm}
\label{th:prime}
Let $\mathcal{C}$ be a quasi-cyclic code of length $pm$ and index $p$ a prime with cyclic constituent codes.
Then the number of quasi-cyclic codes equivalent to $\mathcal{C}$ is $p^r$,
where $r$ is equal to the number of irreducible factors of $Y^m-1$.
\end{thm}
\pf
Under the previous hypotheses, the components $C_i$, $C_j'$ and $C_j''$ of $\mathcal{C}$ are cyclic.
If $\mu_a$ is a multiplier, then the quasi-cyclic code with components $\mu(C_1)$, $C_i$, $i\ne 1$, $C_j'$ and $C_j''$
is equivalent to $\mathcal{C}$.
This also holds for quasi-cyclic codes with components $C_1$, $\mu_a(C_2)$, $C_i$, $i \ne 2$, $C_j'$ and $C_j''$.
It is also true for the quasi-cyclic code with the constituent codes $\mu_a(C_{k})$, $k\in \{1,\le s\}$
or $k \in \{1 \le t \}$ and all others equal to
$C_i$, $C_j'$ or $C_j''$.
Since there are $p-1$ multipliers and $r$ components,
the number of quasi-cyclic codes equivalent to $\mathcal{C}$ which differ in only one component
($\mu_a(C_k)$) is $r(p-1)$, where $r$ is the number of components of $\mathcal{C}$ which is also
the number of factors of $Y^m-1$.
Similarly, the number of equivalent quasi-cyclic codes which differ from $\mathcal{C}$ in only two components ($\mu_a(C_k)$ and $\mu_b(C_{h})$)
is equal to $\binom{r}{2}(p-1)^2$.
Then the total number of quasi-cyclic codes equivalent to $\mathcal{C}$ is equal
\[
\sum_{k=0}^r \binom{r}{k}(p-1)^k=p^r.
\]
\qed

\section{Isodual Quasi-Cyclic Codes}

In this section, conditions are given on the existence of isodual quasi-cyclic codes over $\F_q$.
We start with the following obvious lemma.

 \begin{lem}
 \label{lem:leven}
If there exists an isodual quasi-cyclic code of index $l$, then $l$ must be even.
\end{lem}
\pf
From Theorem~\ref{main:thm}, a condition for the existence of an isodual quasi-cyclic code
is that the constituent codes $C_i$, $1\le i\le s$, are linear isodual codes of length $l$.
This is possible if and only if $l$ is even.
\qed

\begin{rem}
From~Lemma~\ref{lem:leven} if $l=p$ odd, then none of the $p^r$ equivalent codes of the quasi-cyclic code $\mathcal{C}$ of length~$p\cdot m$ given in~Theorem~\ref{th:prime} can be the dual of the code $\mathcal{C}$.
\end{rem}

The results in the remainder of this section are based on the existence of isodual cyclic codes.
Thus we first consider the existence of these codes.

Recall that the multiplier given in (\ref{eq:ling}) is a special kind of permutation which characterizes the
equivalence of some codes.
This multiplier also acts on polynomials of $R[x]$ and thus gives the following ring automorphism
\begin{equation}
\begin{array}{ccl}
\label{eq:ling}
\mu_a:R[x]/(x^n-1) &\longrightarrow &
 R[x]/(x^n-1)\\
 f(x)&\mapsto & \mu_a(f(x)) =
f(x^a).
\end{array}
\end{equation}
If $C$ is a cyclic code generated by $f(x)$, then $\mu_a(C)=\langle f(x^a)\rangle$.
Thus two cyclic codes $C=\langle f(x)\rangle $ and $D = \langle g(x) \rangle $ are multiplier equivalent
if there exists a multiplier $\mu_a$ such that $g(x)=\mu(f(x))=f(x^a)$.
This justifies our previous statement that the concept of multiplier equivalent quasi-cyclic codes is more general
than that given in~\cite{sole3}.

\begin{prop}
\label{prop:equivalent}
Let $C$ be a cyclic code of length $n$ over $\F_{q}$ generated by the polynomial $g(x)$ and $\lambda \in \F_{q}^*$
such that $\lambda^n=1$.
Then the following holds
\begin{enumerate}
\item[(i)] $C$ is equivalent to the cyclic code generated by $g^*(x)$, and
\item[(ii)] $C$ is equivalent to the cyclic code generated by $g(\lambda x)$.
\end{enumerate}
\end{prop}

\pf
\begin{enumerate}
\item [(i)]Consider the multiplier
\begin{equation}
\begin{array}{ccl}
\label{eq:ling2}\mu_{-1}:\mathbb{F}_q[x]/(x^n-1) &\longrightarrow &
 \mathbb{F}_q[x]/(x^n-1)\\
 f(x)&\mapsto & \mu_{-1}(f(x)) = f(x^{-1}),
\end{array}
\end{equation}
which is a ring automorphism.
Assume that deg$(g(x))=r$.
If $C_1$ is the code generated by $g^*(x)$,
then
$C_1=\{ x^rg^{-1}(0)\mu_{-1}(g(x))f(x) \pmod{x^n-1}; f(x)\in\F_q[x]/(x^n-1)\}$.
Clearly $\{ x^rf(x) \pmod{x^n-1}; f(x)\in\F_q[x]/(x^n-1)\}=\{\mu_{-1}(a(x)) \pmod{x^n-1}; a(x)\in\F_q[x]/(x^n-1)\}$, so that
$C_1=\{g(0)^{-1}\mu_{-1}(g(x)a(x)) \pmod{x^n-1}; a(x)\in\F_q[x]/(x^n-1)\}=\mu_{-1}(C)$.
Hence $C$ is equivalent to $C_1$ because $\mu_{-1}$ is a permutation of the coordinates $\{1,x,x^2,\ldots,x^{n-1}\}$.
\item [(ii)] Suppose there exists $\lambda \in \F_{q}^*$ such that $\lambda^n=1$ and let
\[
\begin{array}{cccc}
\phi: \F_{q}[x]/(x^{n}-1)  & \longrightarrow & \F_{q}[x]/(x^{n}-1) \\
                     f(x) & \longmapsto & \phi(f(x))=f(\lambda x). \\
\end{array}
\]
Clearly $\phi$ is a ring automorphism of $\F_q[x]$.
Since $\phi(f(x)+ h(x)( x^{n}-1))=\phi(f(x))+\phi(h(x))(x^n-1)$ as $(\lambda x)^{n}-1=x^{n}-1$,
$\phi$ is well-defined on the ring $\F_{q}[x]/(x^{n}-1)$ and is a ring automorphism of
$\F_{q}[x] /(x^{n}-1)$.
Let $C_2$ be the cyclic code generated by $g(\lambda x)$.
Arguing as in part (i), $C_2=\phi(C)$.
Then because $\phi$ is a diagonal matrix on the coordinates $\{1,x,x^2,\ldots,x^{n-1}\}$,
so that $C$ is equivalent to $C_2$.
\end{enumerate}
\qed
\begin{prop}
\label{lem:decom}
Let $n$ be a positive integer.
If $f(x)$ and $g(x)$ are polynomials in $\F_q[x]$ such that
\begin{equation}
\label{eq:de}
 x^n-1=g(x)f(x),
\end{equation}
then the cyclic code generated by $g(x)$ is equivalent to
the dual of the cyclic code generated by $f(x)$.
\end{prop}

\pf
Let $C_1$ the cyclic code generated by $g(x)$ and $C_2$ the cyclic code generated by $f(x)$.
Since the dual of $C_2$ is generated by $g^*(x)$, by
Proposition~\ref{prop:equivalent}(i) $C_1$ is equivalent to $C_2^\perp$.
\qed

\begin{thm}
\label{thm:equivalent2}
Let $s$ be an odd integer and $f(x)$ a polynomial over $\F_q$ such that $x^s-1=(x-1)f(x)$.
Then the cyclic codes of length $2s$ generated by $(x-1)f(-x)$ and $(x+1)f(x)$ are isodual codes.
\end{thm}

\pf
If $x^s-1=(x-1)f(x)$, then $x^s+1=(x+1)f(-x)$ and
\[
x^{2s}-1 =(x^s-1) (x^s+1)=(x-1)f(x)(x+1)f(-x).
\]
Let $g(x)=(x-1)f(-x)$ be the generator polynomial of a cyclic code $C$.
Then the dual code $C^\perp$ is generated by
\[
h^*(x) =(x+1)f^*(x) = g^*(-x).
\]
Hence from Proposition~\ref{prop:equivalent}(i), $C$ is equivalent to the cyclic code generated by $ g^*(x)$.
Further, from Proposition~\ref{prop:equivalent}(ii), the cyclic code generated by $ g^*(x)$ is equivalent
to the cyclic code generated by $g^*(-x)=h^*(x)$, as the latter code is $C^\perp$,
so that $C$ is isodual.
The same result holds for $g(x)=(x+1)f(x)$.
\qed

\begin{thm}
There exists no self-dual or isodual multiplier quasi-cyclic codes with cyclic constituents over $\F_q$ if $q$ is odd.
When $l=2$, there always exists a quasi-cyclic code with cyclic constituent codes which is isodual.
Further there exists an isodual quasi-cyclic code over $\F_q$ of index $l=2s$ for $s$ odd.
\end{thm}
\pf
Assume the existence of a quasi-cyclic code with cyclic constituents which is also self-dual code, respectively isodual. Hence for $1\i\le s$ the constituent $C_i$ must be cyclic and self-dual, respectively cyclic isodual code that is from~Theorem~\ref{main:thm} and Proposition~\ref{prop:lim}. It is well known that there exists no cyclic self-dual codes cyclic codes~\cite{jia2}, respectively there no cyclic multiplier isodual code if $q$ is odd.
If $l=2$, then $x^2-1=(x-1)(x+1)$, and so from Proposition~\ref{prop:equivalent}(i) the code generated by
$(x-1)$ is equivalent to the code generated by $x+1$, which is its dual.
We consider the quasi-cyclic code with cyclic constituent codes
$C_i =\langle (x-1)f(x) \rangle$ and $C_j'=C_j" =\langle (x-1)f(x) \rangle$.
Since $C_j'=C_j"$ and they are over the same field extension (the degree of $g$ is the same as of $g^*$),
the result follow from Theorem~\ref{main:thm}.
\qed

\end{document}